# Fractional Order Heat Equation in Higher Space-Time Dimensions

Dimple Singh [a, 1], Bhupendra Nath Tiwari [b, 2], Nunu Yadav [c, 3]

[a, b, c] Amity School of Applied Sciences,
Amity University Haryana
Gurgaon, India

[b] INFN-Laboratori Nazionali di Frascati
Via E. Fermi 40, 00044 Frascati, Italy

and

[b] University of Information Science and Technology,
St. Paul the Apostle, Partizanska bb, 6000 Ohrid,
Republic of Macedonia

**Abstract:**

In this paper, we study fractional order heat equation in higher space-time dimensions and offer specific role of heat flows in various fractional dimensions. We offer fractional solutions of the heat equations thus obtained, and examine the associated implications in various limiting cases. We anticipate perspective applications of fractional heat flow solutions in physical systems.

---

[1] dsingh@ggn.amity.edu
[2] bntiwari.iitk@gmail.com
[3] nununonstop006@gmail.com

## 1. Introduction

The partial differential equations whose fractional solutions we consider in this paper are the heat equations. This equation plays a major role in developing laws of physics while studying the real world phenomena occurring at a given scale, including both the macroscopic and microscopic descriptions. In order to focus on the rational order extension of the solutions to ordinary and partial differential equations, we first concentrate on the microscopic versus macroscopic motivations undermining the De Broglie consideration of the wave particle duality.

### 1.1 Heat Equation:

The heat equation is a parabolic partial differential equation that describes the distribution of heat, namely, the variations in the temperature in a given spatial region over time. Given a temperature profile $T(x, t)$, the heat equation in one spatial dimension [1] is given by

$$\frac{\partial T}{\partial t} = K \frac{\partial^2 T}{\partial x^2},$$

where $K$ is a positive constant defining the thermal diffusivity.

The heat equation could be derived from Fourier law and conservation of the energy [1]. Indeed, it is worth mentioning that the heat equation is of fundamental importance in diverse scientific fields ranging from physics, chemistry and mathematics. For instance, in the theory of differential equations, it is a prototype of parabolic partial differential equations [2]. Similarly, in the probability theory, the heat equation is useful in the study of Brownian motion via the Fokker-Planck equation [3]. The importance of the heat equation does not stop here. But, it continues in other domains, as well. For example, in the financial mathematics, the heat equation could be used to solve the Black-Scholes partial differential equation [4]. Physically, the second law of thermodynamics [5] ensures that the heat flows from a hotter region to colder regions.

In this paper, the above condition is analyzed in the context of the fractional calculus, enabling one to study the heat flow with finer details; namely, we can get the differentiation of the temperature profile between given two integers. The above study enables us in high-lightening the importance of fractional order heat equations, where the step sizes of the variables undermining the temperature profile is considered over rational/ real numbers.

In the sequel, we review fundamentals of the fractional calculus in section 2. In section 3, as one of the main theme of the present paper, we focus on the formulation of fractional order partial differential equations, namely, the fractional heat equation. In section 4, we offer respective fractional solutions to the above mentioned fractional differential equations thus developed and their respective generalized solutions using fractional order derivatives. Finally, in section 5, we present conclusions and open problems for future research.

## 2. Review On Fractional Calculus

In this section, we wish recalling basics of fractional calculus. Historically, the fractional calculus is a term used for the theory of derivatives and integrals of arbitrary orders over a rational/ real number, which generalizes the notion of ordinary integer order differentiations and $n$-fold integrations, see Ref. [6] for a review on fractional calculus. The central inspiration behind the fractional calculus is to generalize the standard definitions of differentiation and integration with an integral order n ∈ ℕ to a real order s ∈ ℝ. The first discussion on the fractional calculus began as earlier as 1695 in a letter of Leibniz to L'Hopital addressing about the calculus of an arbitrary order. Hitherto, the fractional calculus is of age three centuries. Amongst the others, Abel, Liouville, Riemann, Euler, and Caputo laid down the foundations of the fractional calculus [7, 8]. Subsequently, it is worth mentioning that the fractional calculus finds various research applications in pure sciences, financial mathematics, applied mathematics, and engineering and technology, e.g., see Ref. [9] for deterministic fractional models in bioengineering and nanotechnology.

Further, it is worth mentioning that the derivatives of non-integer orders are valuably observed in describing physical and chemical properties of various real materials including polymer, rocks and other states of matter [10]. As for as the step size is concerned, the fractional order models were found more logical in revealing the undermining properties lying between the chosen two integers than the corresponding discussions rendering from the respective integer order models. In this paper, we focus on fractional order partial differential equations and their fractional solutions. There are different definitions of fractional derivative and fractional integrals, see Refs. [11, 12] and references therein. Some of definitions of the fractional derivative used in the later sections are enlisted as below.

### 2.1 Grunwald-Letnikov Fractional Derivative:

For a given real valued function $f(t)$ and a rational number $p$, the Grunwald-Letnikov definition of the $p$-th order fractional derivative of $f(t)$ with respect to $t$ is given by

$$_aD_t^p f(t) = \lim_{\substack{h \to 0 \\ nh=t-a}} h^{-p} \sum_{r=0}^{n} (-1)^r \binom{p}{r} f(t-rh),$$

where $h$ is the step size and $a$ is a fixed real number.

### 2.2 Riemann-Liouville Derivative:

Considering the above mentioned real valued function $f(t)$, the Riemann-Liouville definition of the $p$-th order fractional derivative with respect to $t$ is given by

$$_aD_t^p f(t) = (\tfrac{d}{dt})^{m+1} \int_a^t (t-\tau)^{m-p} f(\tau) d\tau, \ m \le p < m+1$$

### 2.3 Caputo's Fractional Derivative:

In the above line of the thought, the Caputo's definition of $\propto$-th order fractional derivative of $f: \mathbb{R} \to \mathbb{R}$ reads as

$$^C D^\alpha f(x) = \frac{1}{\Gamma(\alpha-n)} \int_a^x \frac{f^{(n)}(u)}{(x-u)^{(\alpha-n+1)}} , \ n-1 < \propto < n, \propto \in \mathbb{R}, n \in \mathbb{N}.$$

### 2.4 Euler's Fractional Derivative:

In particular, following the integer order standard derivative of a monomial, the Euler's $\propto$-th order fractional derivative of a monomial, viz. $f(t) = t^\beta$ takes the following form

$$\frac{d^\propto}{dt^\propto}[t^\beta] = D_t^\alpha[t^\beta] = \frac{\Gamma(\beta+1)}{\Gamma(\beta+1-\alpha)} t^{\beta-\alpha}, \alpha \in \mathbb{R}$$

where $\Gamma(r)$ is the standard Gamma function for a given $r \in \mathbb{R}$.

In this paper, we shall focus on the Euler's definition of fractional derivatives while solving the aforementioned fractional differential equations. It is imperative to notice that fractional derivatives satisfy almost all the properties that hold for ordinary derivatives with integral orders. Following the general properties of integral order derivative operator: $D_t^n$, $n \in \mathbb{N}$, we summarize below some straightforwardly verifiable properties [10] concerning the fractional order derivatives for a given pair of real valued functions $f, g: \mathbb{R} \to \mathbb{R}$ as under

- $D_t^\propto[f(t)g(t)] = \sum_{k=0}^{\infty} \binom{\propto}{k} D_t^{\propto-k}[f(t)] D_t^k[g(t)]$, where $\binom{\propto}{k} = \frac{\Gamma(\propto+1)}{\Gamma(k+1)\Gamma(\propto+1-k)}$.
- $D_t^\propto[f(t)C] = \sum_{k=0}^{\infty} \binom{\propto}{k} D_t^{\propto-k}[f(t)] D_t^k[C] = D_t^\propto[f(t)]C$ where $C$ is an arbitrary constant.
- $D_t^\propto[h(t) + g(t)] = \sum_{k=0}^{\infty} \binom{\propto}{k} D_t^{\propto-k}[t^0] D_t^k[h(t) + g(t)] = D_t^\propto[h(t)] + D_t^\propto[g(t)]$.
- $D_t^\propto[h(at)] = a^\propto D_x^\propto[h(x)]$ under the scaling $x = at$.
- $D_t^\propto[t^{-m}] = (-1)^\propto \frac{\Gamma(m+\propto)}{\Gamma(m)} t^{-(m+\propto)}$ for a given $m \in \mathbb{R}$.
- $D_t^{\mu+\nu}[f(t)] = D_t^\mu[D_t^\nu(f(t))] = D_t^\nu[D_t^\mu(f(t))]$ under the composition of $D_t^\nu$ and $D_t^\mu$ on $f(t)$.
- $D_t^{-1}[t^\beta] = \frac{\Gamma(\beta+1)}{\Gamma(\beta+1+1)} t^{\beta+1} = \frac{t^{\beta+1}}{\beta+1}$, where $\beta \in \mathbb{R}$ is corresponding to a negative order derivative.

## 3. Formulation Of Fractional Order Partial Differential Equation in Higher Dimension

In this section, we formulate the fractional order fractional heat equation in higher dimensions extending the respective integral order formulations, see Ref. [1] for a basic review on the above class of second order partial differential equation.

### 3.1 Fractional Heat Equation In Higher Dimensions:

To formulate a fractional version of the heat equation, let's consider two statements as in the standard case, see Ref. [1] for instance, namely, the fact that the heat flows in the direction of decreasing temperature and the rate at which the energy in the form of heat is transferred through an area is proportional to the area and the fractional order temperature gradient normal to the chosen area. Thus, it follows that the heat flux through the area $A$, which is normal to the $x$-axis is given by

$$Q = -kA \frac{\partial^r T}{\partial x^r},$$

where r is a fractional number lying between 0 and 1, viz. we have $0 \leq r \leq 1$. Thus, the energy gained or lost by a body of mass $m$ that undergoes in uniform fractional temperature change $\Delta^r T$ may be expressed as

$$\Delta^r E = cm\Delta^r T = c\rho \Delta^r x \Delta^r y \Delta^r z \Delta^r T$$

where the total mass $m$ contained in the fractional region $(\Delta^r x, \Delta^r y, \Delta^r z)$ is given by $m = \rho \Delta^r x \Delta^r y \Delta^r z$ for a given volume density $\rho$. Thus, the energy flowing into the element through the face in XZ-plane in $\Delta^r t$ is

$$\Delta^r E_{XZ} = Q_{XZ} \Delta^r t = -k\Delta^r x \Delta^r z \Delta^r t \left. \frac{\partial^r T}{\partial y^r} \right|_{x+\frac{\Delta^r x}{2},\ y,\ z+\frac{\Delta^r z}{2}}$$

Similar expressions could indeed be obtained for the fractional order changes in energy from other planes, and thus the total fractional change of order $r$ in the total energy $\Delta^r E$ can readily be evaluated. With this consideration, it hereby follows that we have

$$c\rho \Delta^r x \Delta^r y \Delta^r z \Delta^r T = k\Delta^r x \Delta^r z \Delta^r t \left( \left. \frac{\partial^r T}{\partial y^r} \right|_{x+\frac{\Delta^r x}{2},\ y+\Delta^r y,\ z+\frac{\Delta^r z}{2}} - \left. \frac{\partial^r T}{\partial y^r} \right|_{x+\frac{\Delta^r x}{2},\ y,\ z+\frac{\Delta^r z}{2}} \right) + \dots\dots$$

In the limit of $\Delta^r x \to 0$, $\Delta^r y \to 0$, $\Delta^r t \to 0$, the above equation reduces as

$$\frac{\partial^r T}{\partial t^r} = \frac{k}{c\rho} \lim_{\Delta^r y \to 0} \left[ \frac{\frac{\partial^r T}{\partial y^r} - \frac{\partial^r T}{\partial y^r}}{\Delta^r y} \right] + \cdots$$

Herewith, the factional contributions resulting from various faces of the element in $(3+1)$ dimensions along $(x, y, z)$ spatial directions give the following fractional heat equation

$$\frac{\partial^r T}{\partial t^r} = \frac{k}{c\rho} \nabla^{2r} T,$$

where $\nabla^{2r}$ is the fraction Laplacian operator defined as

$$\nabla^{2r} = \frac{\partial^{2r}}{\partial x^{2r}} + \frac{\partial^{2r}}{\partial y^{2r}} + \frac{\partial^{2r}}{\partial z^{2r}}.$$

## 4. Fractional Solutions

In this section, we offer solutions to fractional order differential equations thus formulated in the previous section, namely, the fractional heat equation in higher dimensions by following the setup of the fractional calculus.

### 4.1 Solution Of Fractional Heat Equation In Higher Dimensions:

For the fractional heat equation as brought out into the consideration in section 3.1, namely, for a given thermal diffusivity $K$, it follows that we have

$$\frac{\partial^r T}{\partial t^r} = K \left[ \frac{\partial^{2r} T}{\partial x^{2r}} + \frac{\partial^{2r} T}{\partial y^{2r}} \right] \text{ and}$$

$$\frac{\partial^r T}{\partial t^r} = K \left[ \frac{\partial^{2r} T}{\partial x^{2r}} + \frac{\partial^{2r} T}{\partial y^{2r}} + \frac{\partial^{2r} T}{\partial z^{2r}} \right]$$

in two and three spatial dimensions respectively. Herewith, by taking different values of $r$, we find that the generalized solution to the above fractional heat equation can be expressed as:

$$T_r(x, y, t) = \frac{1}{\left(\frac{px^{2r}}{2r} + c_1\right)^{\frac{1}{n_1 r}}} \cdot \frac{1}{\left(\frac{py^{2r}}{2r} + c_2\right)^{\frac{1}{n_1 r}}} \cdot r \frac{1}{\left(\frac{Kt^r}{r} + c_3\right)^{\frac{1}{n_2 r}}}$$

and

$$T_r(x, y, z, t) = \frac{1}{\left(\frac{px^{2r}}{2r} + c_1\right)^{\frac{1}{n_1 r}}} \cdot \frac{1}{\left(\frac{py^{2r}}{2r} + c_2\right)^{\frac{1}{n_1 r}}} \cdot \frac{1}{\left(\frac{pz^{2r}}{2r} + c_3\right)^{\frac{1}{n_1 r}}}$$

$$\cdot \frac{1}{\left(\frac{pt^r}{r} + c_4\right)^{\frac{1}{n_2 r}}},$$

where $n_1 r = (1 - 2r)$, $n_2 r = (1 - r)$, $p = \frac{\lambda}{K}$ and $\{c_1, c_2, c_3, c_4\}$ are arbitrary constants. It turns out that the parameter $\lambda$ could further be expressed in terms of $K$, wherefore rendering $\{c_1, c_2, c_3, c_4\}$ as solely independent integration constants.

## 5. Conclusion

In this paper, we consider formulation of fractional order heat equations in higher dimensions and offer the corresponding generalized fractional solutions.

The solutions concerning fractional heat equations as brought out in this paper are anticipated to be of a prototype consideration towards perspective developments of the fractional thermodynamics, temperature profiles with fractional order gradients, fractional fluid dynamics, and Schwartz distributions [13] by considering fractional extensions of the standard elliptic, parabolic and hyperbolic partial differential equations. We leave such considerations open for a future research.